\documentclass[aps,prd,preprint,superscriptaddress,tightenlines,nofootinbib]{revtex4}

\usepackage{graphicx}
\usepackage{dcolumn}
\usepackage{bm}
\usepackage{epsfig}

\def \numups{(5.81\pm0.12)\times 10^{6}}

\def \effone{6.81}
\def \erreffone{\pm 0.07}
\def \efftwo{6.23}
\def \errefftwo{\pm 0.06}

\def \pyieldone{32.6}
\def \pyieldtwo{20.1}
\def \peyo{\asy{6.9}{6.1}}
\def \peyt{\asy{5.8}{5.1}}

\def \rat{0.62}
\def \erat{\asy{0.27}{0.20}}

\def \corrat{0.67}
\def \ecorrat{\asy{0.30}{0.22}}
\def \tot{52.4}
\def \etot{\asy{7.5}{6.9}}

\def \prodone{0.82}
\def \stprodone{\asy{0.17}{0.15}}
\def \sysprodone{\pm 0.06}

\def \prodtwo{0.55}
\def \stprodtwo{\asy{0.16}{0.14}}
\def \sysprodtwo{\pm 0.04}

\def \brone{1.63}
\def \stbrone{\asy{0.35}{0.31}}
\def \sysbrone{\asy{0.16}{0.15}}

\def \brtwo{1.10}
\def \stbrtwo{\asy{0.32}{0.28}}
\def \sysbrtwo{\asy{0.11}{0.10}}

\def \prd#1#2#3{{Phys. Rev. D} {\bf#1}, #2 (#3)}
\def \prl#1#2#3{{Phys. Rev. Lett.} {\bf#1}, #2 (#3)}
\def \plb#1#2#3{{Phys. Lett. B} {\bf #1} #2 (#3)}
\def \nima#1#2#3{{\it Nucl. Instr. Meth.} {\bf A#1} #2 (#3)}
\def \zpc#1#2#3{{\it Zeit. Phys.} C {\bf#1} (#3) #2}
\def \etal{{\it et\,al.}}

\def \asy#1#2{{^{+#1}_{-#2}}}
\newcommand{\plus}{\mbox{$^{+}$}}
\newcommand{\minus}{\mbox{$^{-}$}}

\newcommand{\goesto}{\mbox{$\rightarrow$}}

\newcommand{\lplus}{\mbox{$\ell^+$}}
\newcommand{\lminus}{\mbox{$\ell^-$}}
\newcommand{\dilep}{\mbox{$\lplus\lminus$}}
\newcommand{\bbar}{\mbox{$\bar{b}$} }
\newcommand{\bbbar}{\mbox{$b\bbar$}}  
\newcommand{\upsot}{\mbox{$\Upsilon$}{\rm (1S,2S)}}
\newcommand{\upsi}{\mbox{$\Upsilon$}{\rm (1S)}}
\newcommand{\upsii}{\mbox{$\Upsilon$}{\rm (2S)}}
\newcommand{\upsiii}{\mbox{$\Upsilon$}{\rm (3S)}}
\newcommand{\upsns}{\mbox{$\Upsilon$}{\rm (nS)}}
\newcommand{\upsms}{\mbox{$\Upsilon$}{\rm (mS)}}

\newcommand{\etab}{\mbox{$\eta_b$}}
\newcommand{\chib}{\mbox{$\chi_{b}(1P)$}}
\newcommand{\chibp}{\mbox{$\chi_{b}(2P)$}}

\newcommand{\chibpzero}{\mbox{$\chi_{b0}(2P)$}}
\newcommand{\chibpone}{\mbox{$\chi_{b1}(2P)$}}
\newcommand{\chibptwo}{\mbox{$\chi_{b2}(2P)$}}
\newcommand{\chibpot}{\mbox{$\chi_{b1,2}(2P)$}}
\newcommand{\chibpj}{\mbox{$\chi_{bJ}(2P)$}}
\newcommand{\piz}{\mbox{$\pi$}^{0}}

\newcommand{\pipi}{\mbox{$\pi$}^{+}\mbox{$\pi$}^{-}}
\newcommand{\pizpiz}{\mbox{$\pi$}^{0}\mbox{$\pi$}^{0}}
\newcommand{\mev}{\mbox{ MeV}}
\newcommand{\gev}{\mbox{ GeV}}

\newcommand{\epm}{\mbox{$e^\pm$}}

\begin{document}

\preprint{CLEO CONF 03-06}   
\preprint{LP-121}      

\title{
Observation of the Hadronic Transitions
$\chibpot\goesto\omega\upsi$
}
\thanks{Submitted to XXI International Symposium on Lepton and Photon Interactions at High Energies, August 11-16, 2003, Fermi National Accelerator Laboratory, Batavia, IL}

\author{H.~Severini}
\author{P.~Skubic}
\affiliation{University of Oklahoma, Norman, Oklahoma 73019}
\author{S.A.~Dytman}
\author{J.A.~Mueller}
\author{S.~Nam}
\author{V.~Savinov}
\affiliation{University of Pittsburgh, Pittsburgh, Pennsylvania 15260}
\author{G.~S.~Huang}
\author{D.~H.~Miller}
\author{V.~Pavlunin}
\author{B.~Sanghi}
\author{E.~I.~Shibata}
\author{I.~P.~J.~Shipsey}
\affiliation{Purdue University, West Lafayette, Indiana 47907}
% ...
\collaboration{CLEO Collaboration} 
\noaffiliation


\date{\today}

\begin{abstract} 
The CLEO Collaboration has observed the first hadronic 
transition among bottomonium ($\bbbar$) states
other than the dipion transitions among 
vector states, $\upsns\goesto\pi\pi\upsms$.
In our study of $\upsiii$ decays, 
we find a significant signal for 
$\upsiii\goesto\gamma\omega\upsi$ 
that is consistent with radiative decays $\upsiii\goesto\gamma\chibpot$,
followed by $\chibpot\goesto\omega\upsi$.
The branching ratios we obtain are 
${\cal{B}}(\chibpone\goesto\omega\upsi)$
$=\brone\stbrone\sysbrone\%$
and
${\cal{B}}(\chibptwo\goesto\omega\upsi)$
$=\brtwo\stbrtwo\sysbrtwo\%$,
in which the first error is statistical and  
the second is systematic.
\end{abstract}

\maketitle

\begin{figure}
\includegraphics[width=0.9\linewidth]{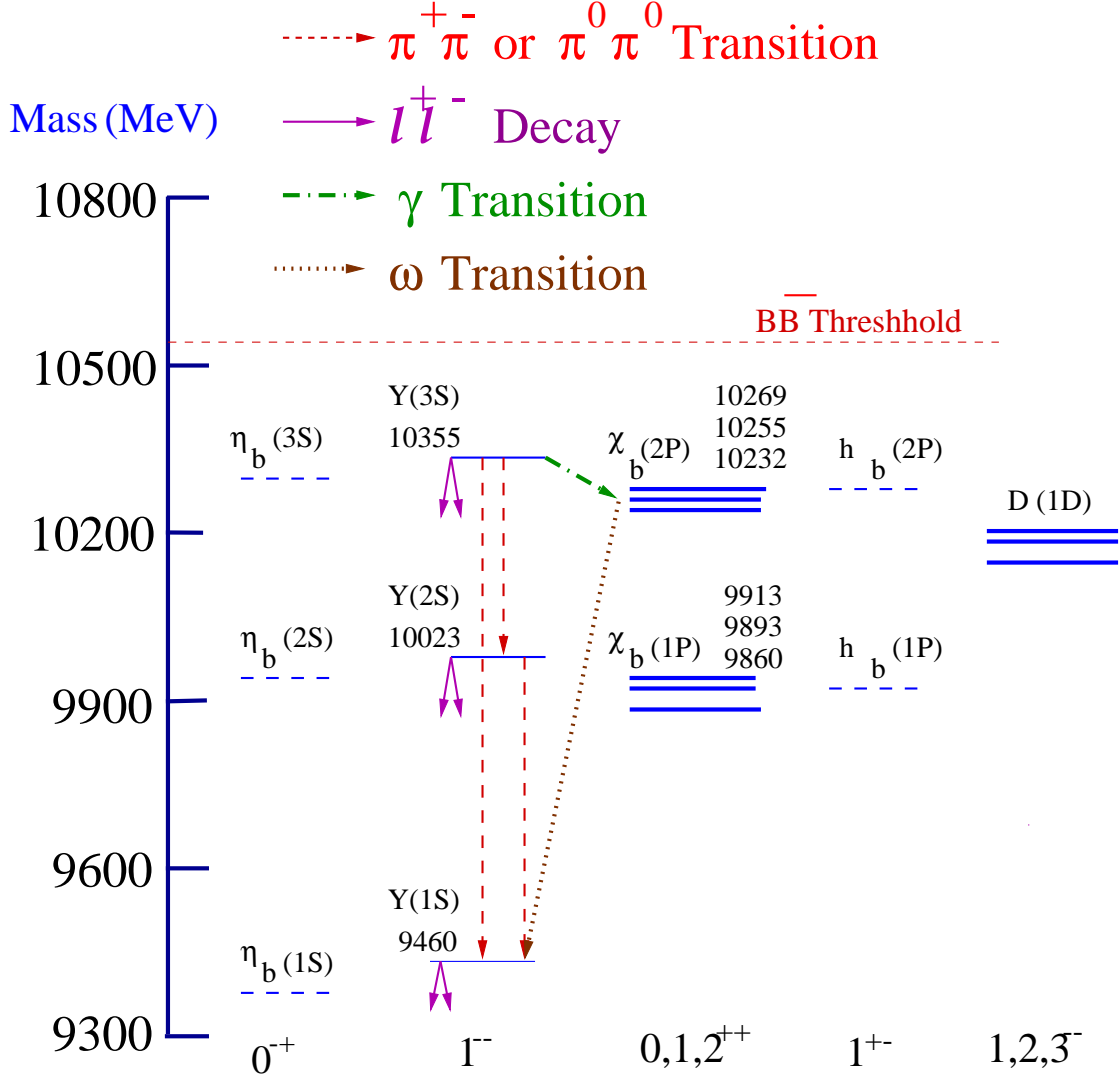}
\caption{The spectrum of bottomonium states. Dilepton decays of 
the vector states are denoted by the solid lines, while dipion
transitions are denoted by dashed lines. The radiative decay
$\upsiii\goesto\chibp$ is indicated by the dot-dashed line, and
the decay $\chibp\goesto\omega\upsi$ by
the dotted line.\label{bbspec}}
\end{figure}

The only hadronic decays of bottomonia that have
been experimentally observed to date are the 
$\pi\pi$ ($\pi\pi\equiv\pipi$ and $\piz\piz$) transitions among
vector bottomonium states.~\cite{expt3s} 
Figure ~\ref{bbspec} illustrates the bottomonium spectrum and 
transitions.
Although there is extensive literature describing hadronic transitions
among bottomonium states~\cite{theory}, nearly all of it 
is directed toward understanding the 
various $\pi\pi$ transitions.~\cite{theory}
While the $\pi\pi$ transitions do provide
important information about strong interaction 
dynamics in heavy quark systems, the investigation
of other hadronic decay modes (i.e. involving $\eta$, 
$\omega$ or multiple $\pi$) 
should offer a different perspective.
The analysis of heavy quarkonium hadronic transitions 
is one of few possible laboratories for the study of 
the physics of the soft gluon emission and 
hadronization process that governs such decays. 
In this letter, we describe the evidence for 
observation of new hadronic transitions, 
$\chibpot\goesto\omega\upsi$. 


The data set analyzed for the purposes of this paper consists of
$\numups\upsiii$ decays observed with 
the CLEO III detector at the Cornell Electron Storage Ring.
The CLEO III detector is described in Ref.~\cite{cleo3det}. 
Charged particle tracking is done by the 47-layer drift 
chamber and a four-layer silicon 
tracker which reside in a 1.5T solenoidal magnetic field.
Photons are detected using an electromagnetic calorimeter (CC)
consisting of 7784 CsI(Tl) crystals distributed in a 
barrel geometry.
The particle-identification capabilities of the CLEO-III
detector are not used in the present analysis.


The final state of $\gamma\pipi\piz\dilep$ 
can be selected in a straightforward manner.  
Events with 
two high momentum charged tracks ($p > 4 \gev$) 
and two or three charged low momentum tracks
($0.12 < p < 0.75 \gev$) 
are considered.
The low momentum tracks 
are required to come from the interaction region
using criteria obtained by studying charged pion tracks 
in a sample of events from the 
kinematically similar decay $\upsii\goesto\pipi\upsi$.

We require events to contain a $\upsi$ candidate
by requiring that the two high momentum tracks in the
event have an invariant mass consistent with 
the $\upsi$ mass $(9.3-9.6)\gev$.  
We make no cuts on the track quality variables for 
the lepton candidate tracks, nor do we
require them to satisfy lepton-identification criteria.
The invariant mass requirement alone 
provides a nearly background free sample,
and imposition of lepton-identification criteria 
only leads to larger systematic errors and reduced
signal efficiency without much improvement in 
signal quality.

We require events to have three or four 
showers in the CC that are not matched to 
charged tracks.  Two of these showers must
form an invariant mass within $3\sigma$ of the nominal 
$\piz$ mass.  In the further analysis, we 
utilize the momentum for this $\piz$ candidate
obtained from a mass-constrained kinematic fit.
In addition to the two showers that correspond to the 
$\piz$, events must contain an isolated photon candidate,
between $50$ and $250$ MeV in energy, that does not 
form an invariant mass within $8 \mev \approx 1.5 \sigma$
of the $\piz$ mass
with any other shower.  Further, the polar angle $(\theta)$ 
of the third shower is restricted to $|\cos{\theta}| < 0.804$, 
the angular region in which CLEO's energy resolution is
sufficiently good to resolve the photons for the signal
channels of interest.  Finally, 
we allow events to contain up to one additional shower 
in the range $|\cos{\theta}|<0.804$.  

In addition, we allow for the possibility of one spurious charged 
track candidate in addition to the four ``signal'' tracks.
Such spurious tracks may arise from failures in pattern recognition
and from delta rays.  Spurious showers may arise from synchrotron 
radiation from the $\epm$ beams or as a result of 
random noise in the CC.
If a given event yields more than one candidate due to
the presence of an additional shower or track,
we choose the candidate for which the sum of energies of
all final state particles is nearest $M(\upsiii)$.

\begin{figure}[t]
\begin{center}
\includegraphics[width=\linewidth]{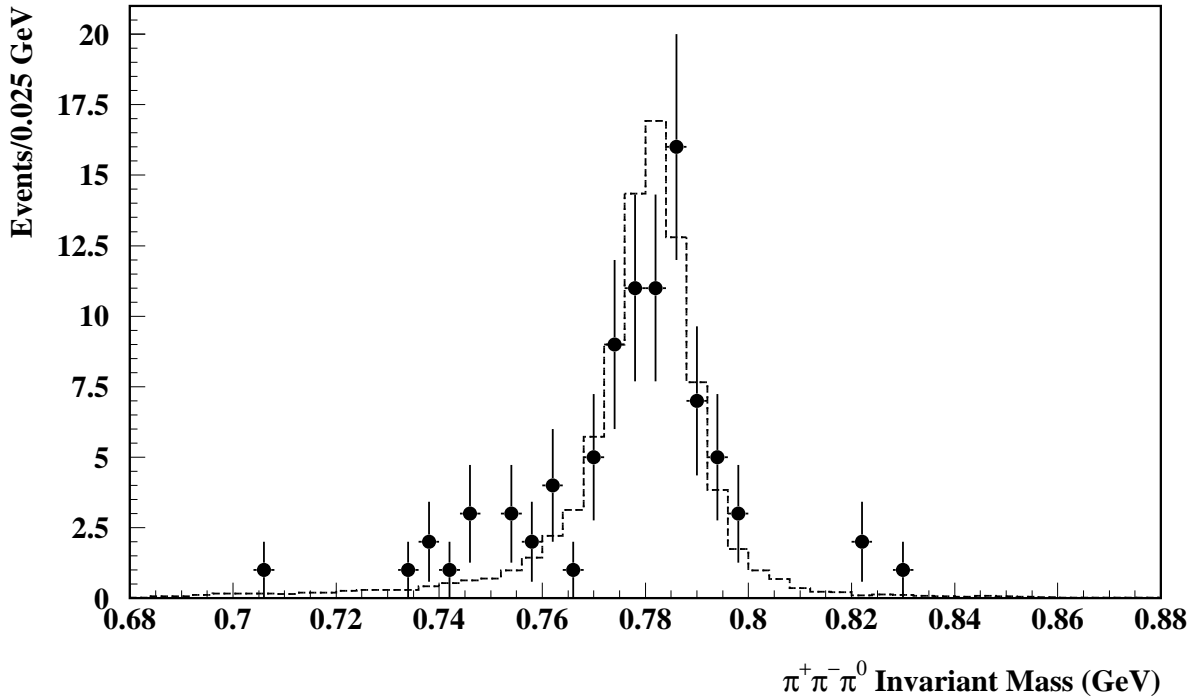}
\end{center}
\caption{$\pipi\piz$ invariant mass, for data events 
subject to final analysis cuts with the exception of 
the cut on the $\chi^2$ of the kinematic fit to $\omega$. \label{3pimass}
The overlaid histogram shows signal Monte Carlo events
(normalized to the same total number) and indicates the 
good reproduction by the Monte Carlo of the 
shape and location of the $\omega$ peak.}
\end{figure}

Because there is no phase space
for a pair of kaons for decays in which a 
$\upsi$ is present,  we assume that the 
low momentum charged tracks are due to $\pi^{\pm}$.
The invariant mass of the $\pipi\piz$ combination,
plotted in Figure~\ref{3pimass},
shows a clear 
enhancement at the mass of $\omega$, 
$M(\omega)=0.782 \gev$.

Finally, to complete the reconstruction of the full
decay chain,
$\upsiii\goesto\gamma\omega\upsi\goesto
\gamma\pipi\piz\dilep$, 
we require that the $\chi^2$ be less than 10 
for a kinematic fit of $\pipi\piz$ constrained 
to the $\omega$ mass,
and subsequently that the
mass recoiling against the
kinematically fitted $\omega$
and $\gamma$ candidates
lie within $\asy{25}{20}$ MeV of $M(\upsi)=9.460 \gev$.  

The simplest explanation for the observed events is the
decay sequence $\upsiii\goesto\gamma\chibpj$, with 
$\chibpj\goesto\omega\upsi$. 
Transitions through $\chibpone$ and $\chibptwo$ only 
are allowed by energy conservation.
$\chibpzero$ lies below threshold for decay to $\omega\upsi$.
In principle a transition through the $\etab(3S)$ state
(see Figure~\ref{bbspec}) is possible, but this state has
never been observed, and, furthermore, the energy of the 
photon in the decay $\upsiii\goesto\gamma\etab(3S)$ is 
expected to be below the range of observed energies in 
the data.  We also neglect as highly improbable the possibility
that the $\omega$ is emitted directly in a transition from
$\upsiii$ to a heretofore unknown state, which then decays
radiatively to $\upsi$.


Backgrounds from ordinary $udsc$ quark production are 
extremely small because of the presence 
of the $\upsi\goesto\dilep$ decay 
in the signal sample.  The only significant source of 
background $\upsi$ expected is known cascades from $\upsiii$.
The final state of $\gamma\pipi\piz\upsi$ may be reached through:
\begin{eqnarray*}
\upsiii&\goesto&\gamma\chibp,\;
\chibp\goesto\gamma\upsii,\;
\upsii\goesto\pipi\upsi\\
\upsiii&\goesto&\piz\piz\upsii,\;
\upsii\goesto\pipi\upsi.
\end{eqnarray*}
In the first case, this final state 
can be produced by 
the addition of a spurious shower in the calorimeter.
In the second case, it 
may be reached by loss of one of the
$\gamma$ from one of the $\piz$ 
due to acceptance or energy threshold.

Backgrounds produced through either of these 
two processes may be removed by excluding events in which
the mass recoiling against the two charged pions 
in the $\upsiii$ reference frame 
is consistent with the hypothesis that the 
$\pipi$ system is recoiling against 
$\upsii$ in the process $\upsiii\goesto X+\upsii
\goesto X+\pipi\upsi$ (between 9.78 and 9.81 GeV).

Two other decay sequences can yield
the final state of 
$\gamma\pipi\piz\upsi$:
the decay $\upsiii\goesto\pipi\upsii$, with
the $\upsii$ decaying either to $\pizpiz\upsi$,
or to $\gamma\chib$ followed by $\chib\goesto\gamma\upsi$.
In each of these cases, however, the $\pi^{\pm}$ are
too soft to produce spurious $\omega$ candidates
for the signal decay chain.

In order to evaluate residual background, we 
generated a GEANT~\cite{geant} Monte Carlo sample for the channel
$\upsiii\goesto\gamma\chibp,\;\chibp\goesto\gamma\upsii,\;
\upsii\goesto\pipi\upsi$,
corresponding to $21.5 \pm 3.1$ million $\upsiii$ decays,
or $4.53\pm0.65$ times our data set.
The uncertainty on the equivalent number of 
$\upsiii$ decays is due to the error on the branching
ratios needed to convert our number of generated events
to the equivalent number of $\upsiii$ decays.
This Monte Carlo sample produced one event
that satisfied our selection criteria.
We therefore expect $1/(4.53\pm 0.65) = 0.22 \pm 0.03$ events 
due to this source.  We also generated a Monte Carlo sample of
$\upsiii\goesto\piz\piz\upsii,\;\upsii\goesto\pipi\upsi$,
corresponding to $538\asy{107}{77}$ million $\upsiii$ decays,
or $113\asy{23}{16}$ times our data set.
From this sample, a total of nine events passed our selection.
In our data set, we thus 
expect $9/(113\pm\asy{23}{16}) = 0.08 \pm 0.01$ events 
due to this source.
To account for the residual background, 
we subtract the expected contribution of
0.30 events from the observed yield. 
We conservatively set a systematic error 
of $\pm 0.30$ events due to this subtraction.

To evaluate the signal detection efficiency, we 
generated 150,000 signal Monte Carlo events 
for each of $\chibpone$ and
$\chibptwo$, proceeding through the sequence
$\upsiii\goesto\gamma\chibpot\goesto
\gamma\omega\upsi\goesto\gamma\pipi\piz\dilep$,
and utilizing PDG average values for the 
$\chibpot$ and $\upsns$ masses, and with a uniform angular 
distributions for the $\upsiii\goesto\gamma\chibpot$ 
and $\upsi\goesto\dilep$ decays.
The analysis cuts described above have been 
applied to these samples, and we obtain
$\epsilon(\chibptwo) = (\efftwo\errefftwo)\%$
and
$\epsilon(\chibpone) = (\effone\erreffone)\%,$
including all selection criteria, acceptance and trigger efficiencies.
The efficiencies have been corrected for the known angular 
distributions of the $E1$ transitions $\upsiii\goesto\gamma\chibpot$. 
We apply an additional relative systematic error of $\asy{0}{3}\%$
to the efficiency in order to account for the possibility that
the $\upsi$ retains the initial polarization of the $\upsiii$.   

\begin{figure}
\includegraphics[width=\linewidth]{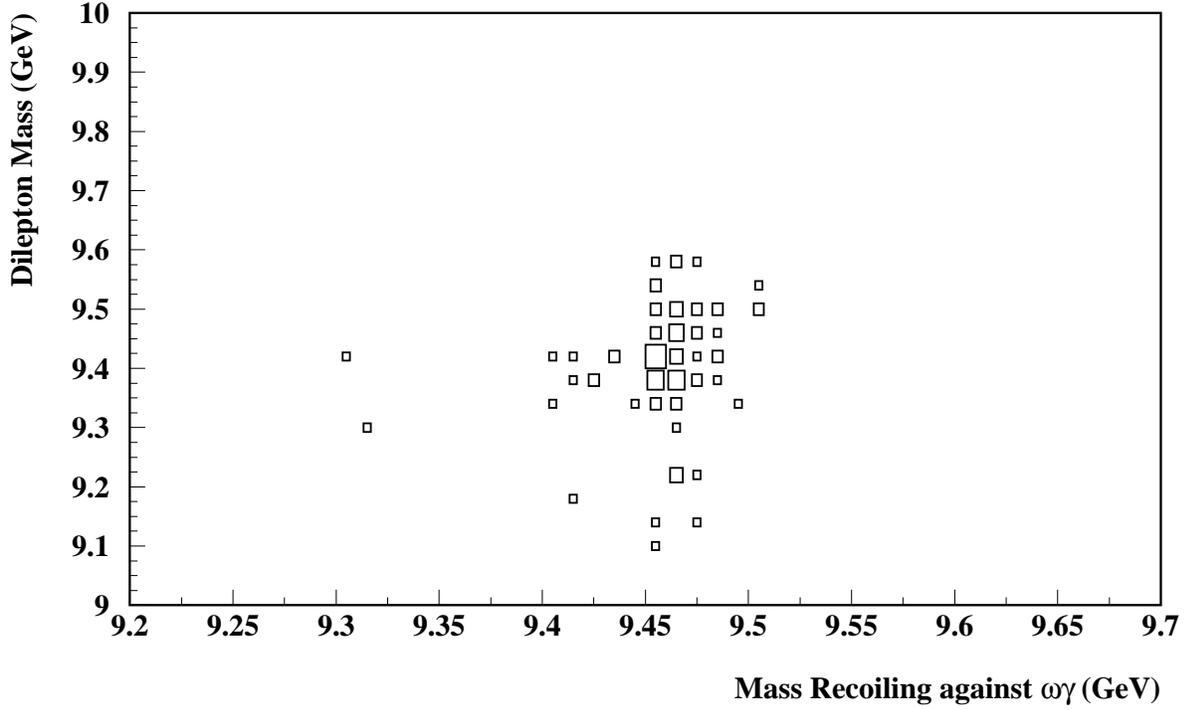}
\caption{Plot of recoil mass versus dilepton mass, for data events
subject to the final set of cuts, with the exception of 
the cuts on the two variables plotted.  
~\label{boxplot}}
\end{figure}

\begin{figure}
\includegraphics[width=\linewidth]{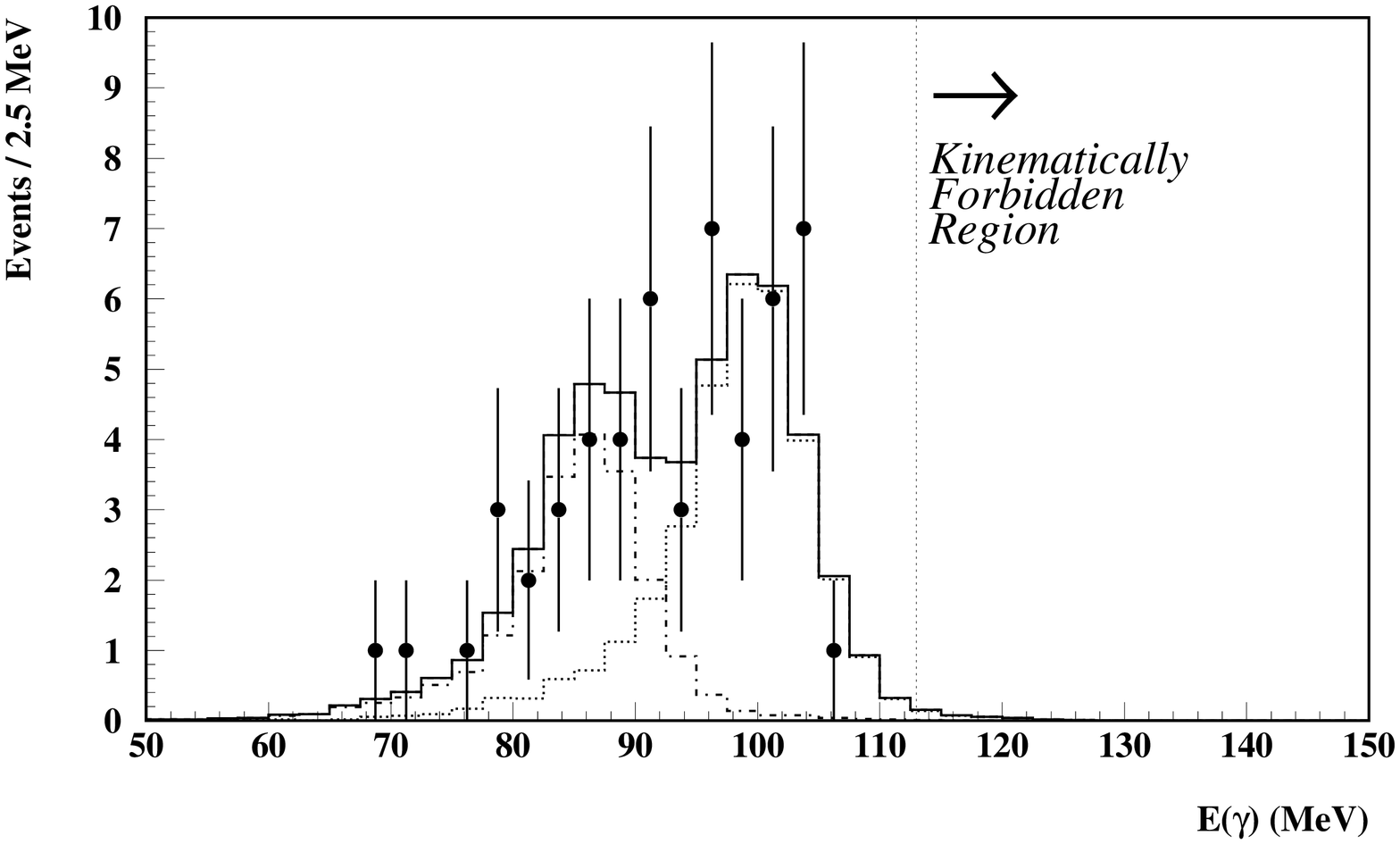}
\caption{Fitted photon energy spectrum for the
final selection of events.  The solid histogram shows contributions
for both $\chibpone$ and $\chibptwo$, while the 
dotted and dashed histograms show the individual $\chibpone$ and
$\chibptwo$ contributions, repsectively.
The dotted line indicates the region above which $\gamma$ 
energies are disallowed for lack of phase space.  
The small leakage of the histogram for the 
$\chibpone$ Monte Carlo sample 
into the kinematically disallowed region 
is due to the finite detector resolution.
~\label{fitboth}}
\end{figure}

In Figure~\ref{boxplot} we present a scatter plot 
of the recoil mass against $\omega\gamma$ vs. dilepton invariant
mass for all events subject to all the cuts discussed above,
except those on the variables plotted,
in order to emphasize the clarity of the signal.  
The final $E_{\gamma}$ spectrum is
shown in Figure ~\ref{fitboth}.
The observed yield has possible contributions from both
decay sequences involving $\chibpone$ and $\chibptwo$ intermediate
states. 

To obtain branching fractions for the 
$\chibptwo$ and $\chibpone$ transitions, 
we perform a maximum likelihood fit of
the $E_{\gamma}$ spectrum.
The expected photon spectra for $\upsiii$ transitions to 
$\chibptwo$ and $\chibpone$ were obtained from 
the signal Monte Carlo samples, and the observed
$E_{\gamma}$ spectrum was then fit to normalized Monte Carlo 
lineshapes with intensities (or yields) for $\chibpone$ and $\chibptwo$ 
as the free parameters. 
We obtain yields of $\pyieldone\peyo$ and $\pyieldtwo\peyt$ events,
respectively. These yields have statistical significances of 
$10.2\sigma$ and $5.2\sigma$, respectively.  
The histogram resulting from the best fit is shown superimposed
on the data in Figure~\ref{fitboth}. 


We subtract the expected background contribution of 
a total of 0.30 events, subtracting from each yield 
a fraction of this total background equal to the 
ratio of the fitted yield to the sum of the two yields. 

We thus obtain the following product branching ratios, using the
detection efficiency and number of $\upsiii$ discussed above:
\begin{eqnarray}
{\cal{B}}(\upsiii\goesto\gamma\chibpone)
\times
{\cal{B}}(\chibpone\goesto\omega\upsi)\nonumber\\
\times
{\cal{B}}(\omega\goesto\pipi\piz)
\times
{\cal{B}}(\upsi\goesto\dilep)\nonumber\\
= \prodone\stprodone\sysprodone\times 10^{-4},\;\mbox{and}\\
{\cal{B}}(\upsiii\goesto\gamma\chibptwo)
\times
{\cal{B}}(\chibptwo\goesto\omega\upsi)\nonumber\\
\times
{\cal{B}}(\omega\goesto\pipi\piz)
\times
{\cal{B}}(\upsi\goesto\dilep)\nonumber\\
= \prodtwo\stprodtwo\sysprodtwo\times 10^{-4},
\end{eqnarray}
in which the first error is statistical and the second is
the systematic error.

The statistical error of nearly $20\%$ is the dominant
source of error.  Systematic error contributions to the above
product of branching ratios are the following:
$2\%$ for the uncertainty in the number of $\upsiii$,
$1\%$ per charged track (a total of $4\%$) for track finding,
$5\%$ for $\piz$ reconstruction, $2\%$ for radiative $\gamma$
reconstruction, $1\%$ for Monte Carlo statistics,
$\asy{3}{0}\%$ for the assumption of uniform $\upsi\goesto\dilep$
angular distribution,
and $1\%$ for background subtraction.  
These contributions, added in quadrature, result in an
overall relative systematic error of 
$\asy{7.7}{7.1}$.

Using the present world average 
branching fractions for 
$\upsiii\goesto\gamma\chibpot$,
$\upsi\goesto\dilep$ (taken to be 
twice that of 
$\upsi\goesto\mu\plus\mu\minus)$,
and $\omega\goesto\pipi\piz$ from Ref.~\cite{pdg2002},
we thus obtain
\begin{eqnarray}
{\cal{B}}
(\chibpone\goesto\omega\upsi)
=\brone\stbrone\sysbrone\%\;\mbox{and}\\
{\cal{B}}
(\chibptwo\goesto\omega\upsi)
=\brtwo\stbrtwo\sysbrtwo\%,
\end{eqnarray}
in which the errors quoted are statistical
and systematic, respectively.
The systematic errors on include the 
experimental errors on
the branching ratios for 
$\upsiii\goesto\gamma\chibpot$, $\omega\goesto\pipi\piz$ 
and $\upsi\goesto\dilep$, 
which 
contribute an additional systematic error of $5.9\%$.

We may also calculate the ratio of $\chibptwo$ to
$\chibpone$ branching ratios, for which systematic
errors discussed above cancel almost entirely.
We obtain this through a maximum likelihood fit
to the $E_{\gamma}$ spectrum, in which the two free
parameters are the sum of yields and the ratio of
$\chibptwo$ to $\chibpone$ yields.
When this fit is performed, we obtain a sum of 
yields equal to 
$\tot\etot$ and a ratio of $\rat\erat$.
In order to convert the ratio of yields 
to the ratio of branching ratios, we 
multiply the yield ratio by 
a factor of $\epsilon(\chibpone)/\epsilon(\chibptwo)$.
\begin{eqnarray}
{\cal{B}}(\chibptwo\goesto\omega\upsi))/
{\cal{B}}(\chibpone\goesto\omega\upsi)
=\rat\erat\times
\frac{\epsilon(\chibpone)}{\epsilon(\chibptwo)}\nonumber\\
=\rat\erat\times
\frac{(\effone\erreffone)\%}{(\efftwo\errefftwo)\%}=\corrat\ecorrat.
\end{eqnarray}

The only systematic errors which do not cancel in this ratio
are the small uncertainties in 
efficiency and 
$\upsiii\goesto\gamma\chibpot$ branching
ratios.  These are entirely negligible compared to
the output error from the maximum likelihood fit.  

In Ref.~\cite{volo}, Voloshin predicts on the basis of
S-wave phase space factors, that the ratio of partial
widths $\Gamma(\chibptwo\goesto\omega\upsi)/\Gamma(\chibpone\goesto\omega\upsi)$ should be approximately 1.4.  The ratio of 
full widths $\Gamma(\chibptwo)/\Gamma(\chibpone)$ 
lies in the range of 1/(1.5-1.25), using world average 
measurements of ${\cal{B}}(\chibpot\goesto\gamma\upsot)$ and 
theoretical predictions for the rates
$\Gamma(\chibpot\goesto\gamma\upsot)$.  
Thus the branching ratios ${\cal{B}}(\chibpot\goesto\omega\upsi)$
are expected to be approximately equal.
Our measurement is rough agreement with this expectation.

Furthermore, Gottfried~\cite{gott} predicted that, due to the
non-Abelian nature of strong interactions, 
$\Delta L=1,\;\Delta S=0$ hadronic transitions ought to
have sizeable widths.  Our observation provides experimental
evidence for his statement.



We have made the observation of the first hadronic decay modes 
involving non-vector $\bbbar$ states, thanks to the large statistical
advantage of the CLEO-III $\bbbar$ data set over all previous
data sets.  The discovery came in the phase-space limited 
channels $\chibpot\goesto\omega\upsi$, and the ratio of the 
measured branching ratios for the two transitions 
are in rough agreement with theoretical expectations based
on S-wave phase space factors. The observation of these decays
supports the prediction that such transitions should 
have significant rates.




We gratefully acknowledge the effort of the CESR staff 
in providing us with
excellent luminosity and running conditions.
This work was supported by 
the National Science Foundation,
the U.S. Department of Energy,
the Research Corporation,
and the 
Texas Advanced Research Program.

\end{document}